\documentclass{JAC2003}
\usepackage{graphicx}
\usepackage{booktabs}

\begin{document}
\title{APOCHROMATIC BEAM TRANSPORT IN DRIFT-QUADRUPOLE 
SYSTEMS
\vspace{-0.5cm}}
\author{V.Balandin\thanks{vladimir.balandin@desy.de}, 
R.Brinkmann, W.Decking, N.Golubeva \\
DESY, Hamburg, Germany}

\maketitle

\begin{abstract}
Though a straight drift-quadrupole system can not be made an achromat, 
there exists an example of a bend-free drift-quadrupole system which can transport 
certain incoming beam ellipses without introducing first-order
chromatic distortions~\cite{MontRugg}. 
In this paper we show that such a possibility is not a rare 
special case, but a general property. For every drift-quadrupole system 
there exists an unique set of Twiss parameters, which will be transported 
through that system without first order chromatic 
distortions. Moreover, we prove that at the same time these Twiss parameters 
minimize the absolute values of the system chromaticities and also bring 
the second order effect of the betatron oscillations on the longitudinal
dynamics to the minimal possible value.
\end{abstract}

\section{INTRODUCTION}

A straight drift-quadrupole system can not be designed in such a way that  
a particle transport through it will not depend on the difference 
in particle energies. Moreover, this dependence can not be removed even 
in first order with respect to the energy deviations.
One way to overcome this problem was developed in the past
and consists in introducing bending of the central trajectory
by a dipole field with subsequent usage of magnetic multipoles at
suitable locations for correction of chromatic aberrations.

However, it often would be desirable to produce achromatic
(i.e. energy independent) focusing in a straight system, without 
involvement of bending magnets. Because, according to the above discussion, 
it is not possible if one considers transport of individual particles
starting from the same initial conditions in the transverse phase space but
with different energies, we will consider the dynamics of a group of particles,
i.e. particle ensembles. It is clear that the dynamics of a particle ensemble
can be sufficiently different from the behavior of a single particle.
For example, the set of particles uniformly distributed on the unit circle in the plane
remains unchanged for an external observer after an arbitrary rotation around
the origin of the coordinate system, while every single particle changes its position.

So, instead of comparing the dynamics of two particles starting 
from the same transverse position but with different energies,
we will compare the results of tracking through the system two
monoenergetic particle ensembles. 
The transverse distributions of the particles within each ensemble are assumed 
to be uncoupled between horizontal and vertical degrees of freedom 
and, for both ensembles, are matched to the same Courant-Snyder 
quadratic forms.
In other words, instead of the dynamics of particles we will study
the propagation of functions, namely Courant-Snyder quadratic forms.

If one is interested only in the lowest order effects and
because the map of a bend-free drift-quadrupole system does not have
second order geometric aberrations, one can equivalently 
look at first order chromatic distortions of the betatron functions
appearing after their transport through the system.
From this point of view there are examples of drift-quadrupole beamlines
for which by appropriate choice of incoming (energy independent) beta
and alpha functions one can remove first (and sometimes even second) 
order chromatic distortion of the exit beta function. 
But, to the author's  knowledge, it is only the paper~\cite{MontRugg}
which explicitly gives an example of the focusing system for which 
first order chromatic distortions at the exit will vanish with
appropriate choice of the entrance Twiss parameters for both,
exit beta and alpha functions.

In this paper we show that such a possibility is not a rare 
special case, but a general property. For every drift-quadrupole system 
(which is not a pure drift space) there exists an unique set of Twiss parameters 
(apochromatic Twiss parameters), which will be transported through that system 
without first order chromatic 
distortions.\footnote{Following~\cite{MontRugg} we have found convenient
to use the term {\it apochromat} for such type of focusing.} 
Moreover, we prove that at the same time these apochromatic Twiss parameters 
minimize the absolute values of the system chromaticities and also bring 
the second order effect of the betatron oscillations on the 
difference of the average bunch path length from the path length of the
reference particle to the minimal possible value (see formula (\ref{LM_2}) below). 
And in the case of a Gaussian beam they also minimize
the effect of the betatron oscillations on the bunch lengthening. 

Note that our interest in the problem of apochromatic beam transport 
was stimulated by the design of the beam distribution and 
transport lines for the European X-Ray Free-Electron Laser (XFEL) 
Facility, in particular by the design of matching sections
and phase shifter of the post-linac collimation system~\cite{XFEL,ColXFEL}.
One of the specifications to the design of this facility is the requirement
for transport lines from linac to undulators to be able to accept
bunches with different energy (up to $\pm1.5\%$ from nominal energy)
and transport them without noticeable deterioration
of beam parameters.
This will allow to fine-tune the FEL wavelength by changing the electron beam energy 
without adjusting magnet strengths and, even more, will allow 
to scan the FEL wavelength within a bunch train by appropriate programming of 
the low level RF system.  

\section{MAPS AND APOCHROMATS}

As usual, we will take the path length along the reference orbit $\,\tau\,$ 
to be the independent variable and will use a complete set 
of symplectic variables
$\mbox{\boldmath $z$} = (x, p_x, y, p_y, \sigma, \varepsilon)$
as particle coordinates~\cite{MaisRipken, My_1}.
Here $\,x,\,y\,$ measure the transverse displacements from the ideal orbit
and $\,p_x,\,p_y\,$ are transverse canonical monenta scaled with 
the constant kinetic momentum of the reference particle $p_0$.
The variables $\,\sigma\,$ and $\,\varepsilon\,$ which describe
longitudinal dynamics are 

\noindent
\begin{eqnarray}
\sigma\,=\,c\,\beta_0 \, (t_0\,-\,t),
\hspace{0.5cm}
\varepsilon\,=\,
({\cal{E}}\,-\,{\cal{E}}_0) \,/\,(\beta_0^2 \,{\cal{E}}_0),
\label{L10_0} 
\end{eqnarray}

\noindent
where $\,{\cal{E}}_0,\,\beta_0\,$ and $\,t_0 = t_0(\tau)\,$
are the energy of the reference particle, its velocity in terms 
of the speed of light $\,c\,$ and its arrival time 
at a certain position $\,\tau$, respectively.

We will represent particle passage through our system
by a symplectic map $\,{\cal M}\,$ that maps the dynamical variables
$\,\mbox{\boldmath $z$}\,$ from the location $\tau = 0$
to the location $\tau = l$.
We will assume that  
$\,\mbox{\boldmath $z$} = \mbox{\boldmath $0$}\,$ is the fixed
point and that the map $\,{\cal M}\,$
can be Taylor expanded in its neighborhood.
Additionally we will assume that the transverse motion 
is dispersion free (always true for the drift-quadrupole systems) 
and uncoupled in linear approximation, which is
a restriction on the form of the six by six symplectic 
matrix $\,M\,$ of the linear part of our map.

Let $\,g_0\,$ be some function of
the variables $\,\mbox{\boldmath $z$}\,$
given at the system entrance.
Then its image $\,g_l\,$ at the system exit
under the action of the map $\,{\cal M}\,$ is given by
the following relation

\noindent
\begin{eqnarray}
\forall \mbox{\boldmath $z$}
\hspace{0.5cm}
g_l(\mbox{\boldmath $z$}) \,=\,
g_0({\cal M}^{-1}(\mbox{\boldmath $z$})),
\label{FP_0}
\end{eqnarray}

\noindent
which symbolically we will write as
$\;g_l \,=\, :{\cal M}:^{-1} g_0$.

Let us consider some Courant-Snyder quadratic forms

\noindent
\begin{eqnarray}
\left\{
\begin{array}{l}
I_x \, = \,
\gamma_x\, x^{2}  + 2 \alpha_x \, x\, p_x 
+ \beta_x \, p_x^{2}\\
I_y \, = \,
\gamma_y\, y^{2} \, + 2 \alpha_y \, y\, p_y 
+ \beta_y \, p_y^{2} 
\end{array}
\right.
\label{IFB_1}
\end{eqnarray}

\noindent
given at the system entrance.
We will say that the map $\,{\cal M}\,$ 
is an $n$-order ($n \geq 2$) apochromat with respect to
the incoming Courant-Snyder quadratic forms $\,I_x\,$ and $\,I_y\,$
if

\noindent
\begin{eqnarray}
:{\cal M}:^{-1} I_{x, y}
\;-\; 
:M:^{-1} I_{x, y} 
\;=\; 
O\left(|\mbox{\boldmath $z$}|^{n + 2}\right).
\label{IFB_3}
\end{eqnarray}
 
\noindent
We will call the Twiss parameters that enter the
Courant-Snyder quadratic forms satisfying (\ref{IFB_3}) 
apochromatic Twiss parameters.

Note that though in this paper we will discuss second-order
apochromats only,
we have the design of a drift-quadrupole
system which is a third-order apochromat and
we will present this design in a separate publication.

Up to any predefined order $n$ the aberrations of the map $\,{\cal M}\,$
can be represented through a Lie factorization as 

\noindent
\begin{eqnarray}
:{\cal M}: \,=_n\,
\exp(:{\cal F}_{n + 1} + \ldots + {\cal F}_3:) :M: ,
\label{IFB_4}
\end{eqnarray}

\noindent
where $=_n$ denotes equality up to order $n$ 
and each of the functions $\,{\cal F}_m\,$
is a homogeneous polynomial of degree $m$
in the variables $\mbox{\boldmath $z$}$.

Using this representation we can state that the map $\,{\cal M}\,$ 
will be $n$-order apochromat with respect to
$\,I_x\,$ and $\,I_y\,$ if, and only if, all homogeneous
polynomials $\,{\cal F}_m\,$ in (\ref{IFB_4}) can be expressed
as functions of $\,I_x,\,I_y\,$ and $\,\varepsilon\,$ only.

\section{EXISTENCE AND UNIQUES OF APOCHROMATIC TWISS PARAMETERS}

The Hamiltonian of the drift-quadrupole system
expanded up to third order 
in the variables $\mbox{\boldmath $z$}$ then
takes the form  
$\,H =_3 H_2 + H_3,\,$ where

\noindent
\begin{eqnarray}
H_2  = (1/2) \big(p_x^2 + p_y^2 + 
\varepsilon^2 / \gamma_0^2\big) + (k_1 /2) \big(x^2 - y^2\big),
\label{Ham2}
\end{eqnarray}

\noindent
\begin{eqnarray}
H_3 = -
(\varepsilon / 2) \big(p_x^2 + p_y^2 + \varepsilon^2 / \gamma_0^2 \big)
\label{Ham3}
\end{eqnarray}

\noindent
and $k_1 = k_1(\tau)\,$ is the quadrupole coefficient.

Let $\,M(\tau)\,$ be a
fundamental matrix solution of the linearized system 
with the elements $r_{mk}$
driven by Hamiltonian $\,H_2$. Then $\,{\cal F}_3\,$
entering formula (\ref{IFB_4}) can be found as

\noindent
\begin{eqnarray}
{\cal F}_3(\mbox{\boldmath $z$}) 
= -\int_0^l H_3(\tau, M(\tau)\mbox{\boldmath $z$} )\, d \tau
\nonumber
\end{eqnarray}

\noindent
\begin{eqnarray}
= -(\varepsilon \,/\, 2)
\cdot
\left(
{\cal Q}_x(x, p_x) \,+\, {\cal Q}_y(y, p_y)
\,-\,l \,\varepsilon^2 \,/\, \gamma_0^2
\right),
\label{MKJ_1}
\end{eqnarray}

\noindent
where ${\cal Q}_x$ and ${\cal Q}_y$ are quadratic forms.

From (\ref{MKJ_1}) one sees that the map of the drift-quadrupole
system does not have second order geometric aberrations
and that the transverse motion still remains uncoupled with
first nonlinear correction terms taken into account. 
So we will restrict our further consideration to the
motion in one degree of freedom (horizontal).
Let us denote

\noindent
\begin{eqnarray}
{\cal Q}_x(x, p_x) =
c_{20} \,x^2 \,+\, 2 c_{11} \,x p_x \,+\, c_{02} \,p_x^2 .
\label{MKJ_2}
\end{eqnarray}

\noindent
The coefficients of this form are given by the integrals

\noindent
\begin{eqnarray}
c_{20} = -\int_0^l r_{21}^2(\tau) \,d \tau,
\hspace{0.4cm}
c_{02} = -\int_0^l r_{22}^2(\tau) \,d \tau,
\label{MKJ_3}
\end{eqnarray}

\noindent
\begin{eqnarray}
c_{11} = -\int_0^l r_{21}(\tau)\,r_{22}(\tau) \,d \tau
\label{MKJ_4}
\end{eqnarray}

\noindent
and therefore satisfy Cauchy-Bunyakovsky inequality

\noindent
\begin{eqnarray}
c_{20} \,c_{02} \,-\,c_{11}^2 \,\geq\, 0, 
\label{MKJ_5}
\end{eqnarray}

\noindent
where, as it is not difficult to prove, the equality holds if,
and only if, $\,k_1(\tau) \equiv 0$, 
i.e. if our system is the pure drift space.
So if $\,k_1(\tau) \not\equiv 0\,$ then the
quadratic form ${\cal Q}_x$ is negative-definite
and therefore, according to the statement at the
end of the previous section, there exist unique
apochromatic Twiss parameters given by the
following expressions

\noindent
\begin{eqnarray}
\beta_x = -\frac{c_{02}}{\sqrt{c_{20} c_{02} -c_{11}^2}},
\;\;
\alpha_x = -\frac{c_{11}}{\sqrt{c_{20} c_{02} -c_{11}^2}}.
\label{MKJ_8}
\end{eqnarray}

\section{CHROMATIC VARIABLES}

The quantities which are usually used in accelerator physics for the description 
of the first-order chromatic effects
are chromaticity and 
two betatron amplitude difference functions~\cite{MAD}.
It is intuitively clear that all of them
must be connected with each other and with the chromatic map 
aberrations. Unfortunately the usual ways 
of their introduction are formally different and do not allow 
immediately to see such a connection. In this section we introduce 
chromatic variables as coefficients of the expansion of the quadratic form (\ref{MKJ_2})
with respect to three independent quadratic invariants of
linear motion. One of these variables coincides with the usual
chromaticity and two others (apochromaticities) can be expressed through
betatron amplitude difference functions using rotation
and scaling. 

\subsection{Invariants of Uncoupled Linear Motion}

It is well known that for the linear dynamics the Courant-Snyder quadratic
form is an invariant of motion. It is much less known that
in the linear case there are other invariants which depend
on betatron phase and together with the Courant-Snyder invariant
form a basis in the space of polynomials~\cite{My_2}.
For the purpose of this paper we need only two of them
given by the following formulas

\noindent
\begin{eqnarray}
\left\{
\begin{array}{l}
U_x \, = \,\bar{U}_x \cos(2 \mu_x) - \bar{V}_x \sin(2 \mu_x)\\
V_x \,\, = \,\bar{U}_x \sin(2 \mu_x) + \bar{V}_x \cos(2 \mu_x)
\end{array}
\right.
\label{QQ_1}
\end{eqnarray}

\noindent
where

\noindent
\begin{eqnarray}
\bar{U}_x = (2 /\beta_x)\, x^2 - I_x,
\hspace{0.2cm}
\bar{V}_x = 2\big((\alpha_x / \beta_x) \,x^ 2 + x p_x\big).
\label{QQ_3}
\end{eqnarray}

\subsection{Chromaticity and Apochromaticities}

We define chromaticity $\,\xi_x\,$ and apochromaticities
$\,\zeta_x\,$ and $\,\eta_x\,$ as coefficients in the representation

\noindent
\begin{eqnarray}
{\cal Q}_x(x, p_x) =
\xi_x \, I_x(0) \,+\,
\zeta_x \, U_x(0) \,+\,
\eta_x \, V_x(0).
\label{DD_1}
\end{eqnarray}

\noindent
Comparing (\ref{DD_1}) with (\ref{MKJ_2}) we obtain 

\noindent
\begin{eqnarray}
\left\{
\begin{array}{l}
\xi_x =
(1 / 2)
(\beta_x(0) c_{20} - 2 \alpha_x(0) c_{11} + \gamma_x(0) c_{02})\\
\zeta_x = \xi_x - c_{02} / \beta_x(0)\\
\eta_x =
c_{11} - (\alpha_x(0) / \beta_x(0))\,c_{02}
\end{array}
\right.
\label{DD_4}
\end{eqnarray}

From these formulas it is not difficult to derive the
following important equality

\noindent
\begin{eqnarray}
\xi_x^2\,=\,
\zeta_x^2 \,+\, \eta_x^2 \,+\,
\left(c_{20}\, c_{02} \,-\, c_{11}^2 \right).
\label{DD_5}
\end{eqnarray}

With the representation (\ref{DD_1}) the image of the
Courant-Snyder quadratic form after passage through
the system is

\noindent
\begin{eqnarray}
:{\cal M}:^{-1} I_x(0)
=_3
I_x(l) + 2 \varepsilon
\left(\zeta_x V_x(l)-\eta_x \,U_x(l)\right),
\label{QQ_3_4}
\end{eqnarray}

\noindent
and one sees that this quadratic form
will be apochromatic if, and only if, 
$\,\zeta_x=\eta_x=0$. This together with (\ref{DD_5})
tells us that apochromatic incoming
Twiss parameters simultaneously are the Twiss parameters
which minimize the absolute value of the system chromaticity.

Note that for drift-quadrupole systems 
chromatic variables can be represented also
in the form of the integrals

\noindent
\begin{eqnarray}
\xi_x
\,=\, 
- \frac{1}{2} \int_0^{l} \gamma_x \, d \tau,
\label{CandA_1}
\end{eqnarray}

\noindent
\begin{eqnarray}
\zeta_x 
=
\frac{1}{2} \int_0^{l}
\left[
\frac{1 - \alpha_x^2}{\beta_x} \cos(2 \mu_x)
-\frac{2 \alpha_x}{\beta_x} \sin(2 \mu_x)
\right] d \tau,
\label{CandA_2}
\end{eqnarray}

\noindent
\begin{eqnarray}
\eta_x
=
\frac{1}{2} \int_0^{l}
\left[
\frac{1 - \alpha_x^2}{\beta_x} \sin(2 \mu_x)
+ \frac{2 \alpha_x}{\beta_x} \cos(2 \mu_x)
\right] d \tau.
\label{CandA_3}
\end{eqnarray}

\section{EFFECT OF BETATRON OSCILLATIONS ON LONGITUDINAL MOTION}

To see more clearly the effect of the betatron oscillations on the longitudinal
motion, let us assume that the beam at the system entrance is monoenergetic with
$\varepsilon = 0$ for all particles.
If the beam distribution is uncoupled 
between the degrees of freedom
then

\noindent
\begin{eqnarray}
\big<\sigma(l)\big> \,=_2\,\big<\sigma(0)\big> \,+\, 
\epsilon_x \, \xi_x  
\,+\,\epsilon_y\,\xi_y,
\label{LM_2}
\end{eqnarray}

\noindent
where $\epsilon_x$ and $\epsilon_y$ are the non-normalized rms
emittances and the chromaticities $\xi_x$ and $\xi_y$ are calculated
using the beam Twiss parameters.
If, additionally, we will assume that the particle distributions
in each transverse plane are Gaussian, then we can also obtain
the following formula for the effect of the betatron oscillations on
the bunch lengthening

\noindent
\begin{eqnarray}
\Big<
\big[
\sigma(l) - \big< \sigma(l) \big>
\big]^2
\Big>
 =_4
\Big<
\big[
\sigma(0) - \big< \sigma(0) \big>
\big]^2
\Big>
\nonumber
\end{eqnarray}

\noindent
\begin{eqnarray}
+\,
\epsilon_x^2 \cdot \big( \xi_x^2 \,+\, \zeta_x^2 \,+\, \eta_x^2 \big)  
\,+\,
\epsilon_y^2 \cdot \big( \xi_y^2 \,+\, \zeta_y^2 \,+\, \eta_y^2 \big).
\label{LM_3}
\end{eqnarray}

\noindent
So one sees that the effect of the betatron oscillations on the longitudinal
motion becomes minimal for the beam matched to the apochromatic Twiss
parameters.

\end{document}